\documentclass[twocolumn,showpacs]{revtex4}
\usepackage{epsfig,graphicx,amssymb,color,bm}

\begin{document}

\title{Controlling photon transport in the single-photon weak-coupling
regime of cavity optomechanics}
\author{Wen-Zhao Zhang, Jiong Cheng, Jing-Yi Liu and Ling Zhou\footnote{%
Email: zhlhxn@dlut.edu.cn}}

\begin{abstract}
We study the photon statistics properties of few-photon transport in an optomechanical system where an
optomechanical cavity couples to two empty cavities. By analytically deriving the one- and two-photon currents
in terms of a zero-time-delayed two-order correlation function, we show that a photon blockade can be achieved in
both the single-photon strong-coupling regime and the single-photon weak-coupling regime due to the nonlinear
interacting and multipath interference. Furthermore, our systems can be applied as a quantum optical diode,
a single-photon source, and a quantum optical capacitor. It is shown that this the photon transport controlling
devices based on photon antibunching does not require the stringent single-photon strong-coupling condition.
Our results provide a promising platform for the coherent manipulation of optomechanics, which has potential
 applications for quantum information processing and quantum circuit realization.
\end{abstract}


\address{School of Physics and Optoelectronic Technology, Dalian University of
Technology, Dalian 116024,PR China}
\maketitle
\section{introduction}

The nonlinear effect is potential resource for quantum information
processing \cite{QIP}. For example, the photon blockade resulting from the
nonlinearity is employed in single-photon (few-photon) transmission control %
\cite{TC} and optical state truncation \cite{ST}. Similarly, photon blockade
is also an important feature in a lot of quantum device design such as fast
two-qubit controlled-NOT gate \cite{CNOT}, efficient quantum repeaters \cite%
{EQR}, single-photon transistor \cite{SPT} and optical quantum computer \cite%
{QC}. The rectifying device related with nonlinearity is the key device to
information processing in integrated circuits \cite{ED}. Considerable
efforts has been made to investigate the optical diodes \cite{OD}. Recently,
various possible solid-state optical diodes have been proposed, for example
the diodes from standard bulk Faraday rotators \cite{FR}, integrated on a
chip \cite{CP}, realized in opto-acoustic fiber \cite{OF} and from moving
photonic crystal \cite{MPC}. A kind of optical diode based on photon
blockade effect also have been proposed, including photonic diode by a
nonlinear-linear junction of coupled resonators \cite{OPD} and optical diode
of two semiconductor microcavities coupled via $\chi ^{(2)}$ nonlinearities %
\cite{MOD}.

The nonlinear interaction between optical and mechanical modes arising from
radiation pressure force in optomechanical (OM) systems exhibit a lot of
interesting nonlinear effects such as photon (phonon) blockade \cite{PB1,PB2}%
, optomechanical induced transparency \cite{OMIT1,OMIT2} and Kerr
nonlinearity \cite{Kerr1,Kerr2}. Cavity optomechanics has received
significant attention both in fundamental experiments \cite{FE1,FE2} and
sensing applications \cite{SD1,SD2}. Currently, experimental technique of
cavity optomechanics are still in the single-photon weak coupling regime %
\cite{RVOM} ($g^{2}<\kappa \omega _{m}$), meanwhile it draws relatively few
of works as control devices in quantum information processing because the
prerequisite of the strong nonlinear is required \cite{OMD1,OMD2}. In order
to utilize the nonlinearity of OM system in quantum information control,
much attention has been paid to the photon blockade in OM system, including
quadratically coupled OM systems \cite{QOM}, hybrid electro-optomechanical
system \cite{HOM}, and ultrastrong optomechanics \cite{UOM}, where the
strong coupling condition is required. Ref. \cite{ABPB2} has shown that
strong photon antibunching can be achieved in two coupled cavities with weak
Kerr nonlinearity, which motivate us try to achieve strong nonlinear effect
in OM system in weak coupling regime. In this paper, we propose a scheme to
realize an optical diode with optomechanical cavity coupled to two cavities.
This scheme does not require the stringent condition that the single-photon
optomechanical coupling strength $g$ is on the order of the mechanical
resonance frequency $\omega _{m}$ \cite{PB2} or the coupling strength $g$ is
larger than the cavity decay rate $\kappa $ \cite{QOM}. Our results show
that photon blockade can be achieved both in strong and weak coupling regime
because of the nonlinearity and the multipath interference. By examining the
second-order correlation function, rectifying factor $\mathcal{R}$ and
transport efficiency $\mathcal{T}$, we exhibit the characteristics of our
system as photonic diode. Meanwhile, the single-photon transport can be
controlled by the tuning of the frequency of the cavities in strong coupling
regime. Surprisingly, by pumping two sides of the system (cavity $L$ and $R$%
), the device will embodies some characteristics like capacitor: photon
storage-release (charge-discharge) and filtering of photon frequency.

The paper is organized as follows. In Sec. II, we introduce the system and
eigensystem of Hamiltonian. We discuss photons transport control in the
cavity optomechanical system as diode and capacitor in Sec. III. Discussion
and conclusions are given in Sec. IV.

\section{Model and Hamiltonian}

\begin{figure}[tbp]
\centering \includegraphics[width=7cm]{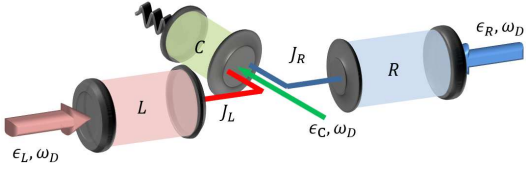}\newline
\caption{(Color online). Schematic of the cavity optomechanical system
coupled with two cavities. Cavities $L$, $R$ and $C$ all can driven by the
laser field with same frequency.}
\label{system}
\end{figure}

We consider a compound optomechanical system in which a cavity with a
movable mirror is coupled with two cavities (L and R) with the coupling
constant $J_{L}$ and $J_{R}$, see Fig. \ref{system}. The system described by
the Hamiltonian $H=H_{sys}+H_{pump}$. By setting $\hbar =1$, the Hamiltonian
$H_{sys}$ reads
\begin{eqnarray}
H_{sys} &=&\sum_{j=L,C,R}\omega _{l}a_{j}^{\dagger }a_{j}+\omega
_{m}b^{\dagger }b+g(b^{\dagger }+b)a_{C}^{\dagger }a_{C}  \nonumber \\
&&+(J_{L}a_{L}^{\dagger }a_{C}+J_{R}a_{R}^{\dagger }a_{C}+h.c.),
\end{eqnarray}%
where $a_C$, $a_{L}$ and $a_{R}$ are the annihilation operator for the
photon mode of cavities $C$, $L$ and $R$ with frequency $\omega _{C}$, $%
\omega _{L}$ and $\omega _{R}$, respectively. $b$ is the phonon annihilation
operator of the mechanical mode for the mirror with frequency $\omega _{m}$,
$g$ denote the coupling strength of radiation pressure. The cavity modes are
driven by the laser with the same frequency $\omega _{D}$, which can be
described by $H_{pump}=\sum_{j=L,C,R}\varepsilon _{j}(a_{j}^{\dagger
}e^{-i\omega _{D}t}+h.c.)$. In the rotating frame with $H_{0}=\sum_{j=L,C,R}%
\omega _{D}a_{j}^{\dagger }a_{j}$, we obtain
\begin{eqnarray}
H_{S} &=&\sum_{j=L,C,R}\Delta _{j}a_{j}^{\dagger }a_{j}+\omega
_{m}b^{\dagger }b+g(b^{\dagger }+b)a_{C}^{\dagger }a_{C}  \nonumber \\
&&+(J_{L}a_{L}^{\dagger }a_{C}+J_{R}a_{R}^{\dagger }a_{C}+h.c.)  \nonumber \\
&&+\sum_{j=L,C,R}\varepsilon _{j}(a_{j}^{\dagger }+a_{j}),
\end{eqnarray}%
where $\Delta _{j}=\omega _{j}-\omega _{D}$ $(j=L,C,R)$ are the detuning
between the driving field and the $jth$ cavity frequency, respectively. For
this cascade configuration, cavity $\emph{L}$ and $\emph{R}$ are used as
input and output ports in the side $\emph{L}$ and $\emph{R}$. In this case,
optomechanical cavity as an assisted-cavity, provides an intrinsically
nonlinear interaction.

We assume that the cavities ($L$ and $R$) incoherently dissipate at rates $%
\kappa _{l}$ $(l=L,R)$ determined by the openness of the output channels and
only classical driving fields are added to the quantum vacuum of the system,
then according to the standard input-output relation \cite{ITR}, the average
output current (or photon stream) as number of quanta emitted at time $t$
from each cavity can be formally given by
\begin{equation}
Q_{l}(t)=\kappa _{l}\emph{Tr}[a_{l}^{\dag }a_{l}\rho (t)],(l=L.R),
\label{output}
\end{equation}%
where $\rho $ is the density operator of system. The evolution of the
density operator $\rho $ for the Hamiltonian $H_{S}$ can be described by the
master equation
\begin{eqnarray}
\dot{\rho} &=&-i[H_{S},\rho ]+\sum_{j=L,C,R}\frac{\kappa _{j}}{2}\mathcal{D}%
[a_{j}]\rho +\frac{\gamma }{2}(n_{th}+1)\mathcal{D}[b]\rho  \nonumber \\
&&+\frac{\gamma }{2}n_{th}\mathcal{D}[b^{\dagger }]\rho ,  \label{Meq}
\end{eqnarray}%
where $\kappa _{j}$ and $\gamma $ are the cavity and mechanical energy decay
rates, $n_{th}=[exp(\omega _{m}/k_{B}T_{M}-1)]^{-1}$ is the average thermal
occupancy number of the oscillator. $\mathcal{D}[o]=2o\rho o^{\dagger
}-o^{\dagger }o\rho -\rho o^{\dagger }o$ is the Lindblad dissipation
superoperator.

The eigen equation of the Hamiltonian $H_{om}=\omega _{m}b^{\dagger
}b+\Delta _{C}a^{\dagger }a+g(b^{\dagger }+b)a^{\dagger }a$ can be expressed
as
\[
H_{om}|s\rangle _{C}|\tilde{n}(s)\rangle _{m}=E_{s,n}|s\rangle _{C}|\tilde{n}%
(s)\rangle _{m},
\]%
where the eigenvalues are
\[
E_{s,n}=s\Delta _{C}+n\omega _{m}-s^{2}\delta ,
\]%
with $\delta =\frac{g^{2}}{\omega _{m}}$, and the eigenstate
\begin{equation}
|\tilde{n}(s)\rangle =e^{g(b-b^{\dagger })/\omega _{m}}|n\rangle
\label{dpns}
\end{equation}%
\begin{figure}[tbp]
\centering \includegraphics[width=8cm]{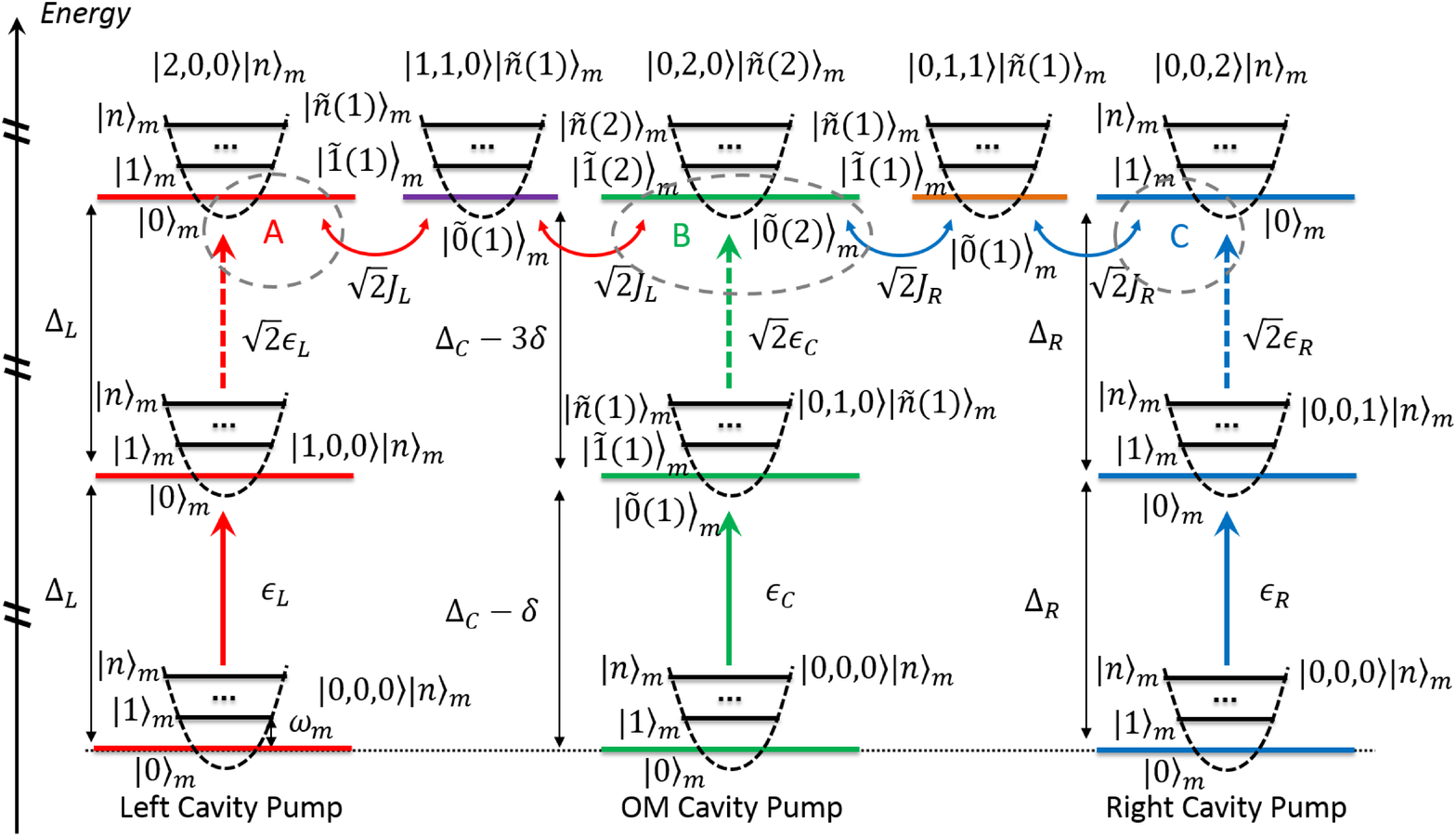}\newline
\caption{(Color Online) The eigensystem of the Hamiltonian $H_{S}$ in the
zero-, one-, and two-photon cases, sub-area $A$, $B$ and $C$ denote
multi-path interference in the system}
\label{egs}
\end{figure}
is the displaced number state. The eigensystem of the Hamiltonian $H_{S}$ in
the zero-, one-, and two-photon cases is shown in Fig. \ref{egs}. We noticed
that, the energy levels for optomechanical cavity (middle green line) will
obtain a shift $s^{2}\delta $ caused by the nonlinear interacting with a
frequency red (blue) detuning from the resonator resonance. This nonlinear
shift can lead to bunched or antibunched photons in the OM cavity (the
details are given in Sec. III C). This nonlinear effects also can appear in
other cavities because of the coupling $J_{L}$ and $J_{R}$. Especially in
strong coupling regime $g/\kappa \gg 1$, the system appears photon blockade,
i.e. the probability for two photons inside the cavity is largely suppressed
due to the energy restriction.

The interference between multipath for two-photon excitation in cavities are
partially responsible for the photon antibunching effect shown in the
sub-area of eigensystem diagram A, B and C. For area A, the two-photon in
cavity $L$ with state $|2,0,0\rangle |n\rangle _{m}$ have two excitation
path, one is direct excitation from low level in cavity $L$ with state $%
|1,0,0\rangle |n\rangle _{m}$, the other is the tunnelling from OM cavity to
left cavity with state $|1,1,0\rangle |\tilde{n}\rangle _{m}$. The
destructive interference between the two paths reduces the probability of
two-photon excitation in the cavity. As well as area B and C. When the
probability equal to zero, unconventional photon blockade \cite%
{ABPB1,ABPB2,UPB} appears in the cavity with no requirement to strong
nonlinear coupling coefficient $g$ (even $g/\kappa <1$). Therefore, the
compound optomechanical system can work as a single photon control device
both in OM weak- and strong- coupling regime, which will be discussed in
detail in next section.

\section{Photons transport control in the cavity optomechanical system}

\subsection{Optomechanical optical diode}

When the nonlinear effect for the right-going ($k$) is different from that
for the left-going ($-k$) waves, i.e. the nonlinearity of the composite
system is asymmetric, the rectification of one-dimensional photons transport
can be controlled. The one way transport is called optical diode \cite{OD}.
In this section, we will show that our compound system can worked as a
photonic diode.

We are interested in the statistic property of photons and its control.
Usually, the frequency of the mechanical oscillator is larger than the
strength of coupling of the radiation pressure, i.e., $\omega _{m}\gg g$.
For simplicity, we can adiabatically eliminating the degree of the
oscillators. Including the decay rate of the cavities, we have non-Hermitian
effective Hamiltonian as
\begin{eqnarray}
H_{eff} &=&\sum_{j=L,C,R}[(\Delta _{j}-i\kappa _{j}/2)a_{j}^{\dag
}a_{j}+\varepsilon _{j}(a_{j}^{\dag }+a_{j})]  \nonumber \\
&&-i\sum_{j=L,C,R}\frac{\kappa _{j}}{2}a_{j}^{\dagger }a_{j}-\delta
a_{C}^{\dag }a_{C}-\delta a_{C}^{\dag }a_{C}^{\dag }a_{C}a_{C}  \nonumber \\
&&+(J_{L}a_{L}^{\dag }a_{C}+J_{R}a_{R}^{\dag }a_{C}+h.c.).  \label{Heff}
\end{eqnarray}%
We assume that the general state is
\begin{eqnarray}
|\psi (t)\rangle  &=&C_{0}(t)|{\o }\rangle
+\sum_{j=L,C,R}C_{j}(t)a_{j}^{\dag }|{\o }\rangle   \nonumber \\
&&+\sum_{i,j=L,C,R}\frac{1}{2}C_{ij}(t)a_{i}^{\dag }a_{j}^{\dag }|{\o }%
\rangle .
\end{eqnarray}%
{Under weak pumping conditions, we have \cite{ABPB2}
\[
C_{0}\gg C_{j}\gg C_{ij}.
\]%
Under this condition, one can obtain the steady state solution of the
probability amplitudes}, see Appendix.

\begin{figure}[tbp]
\centering \includegraphics[width=4.3cm]{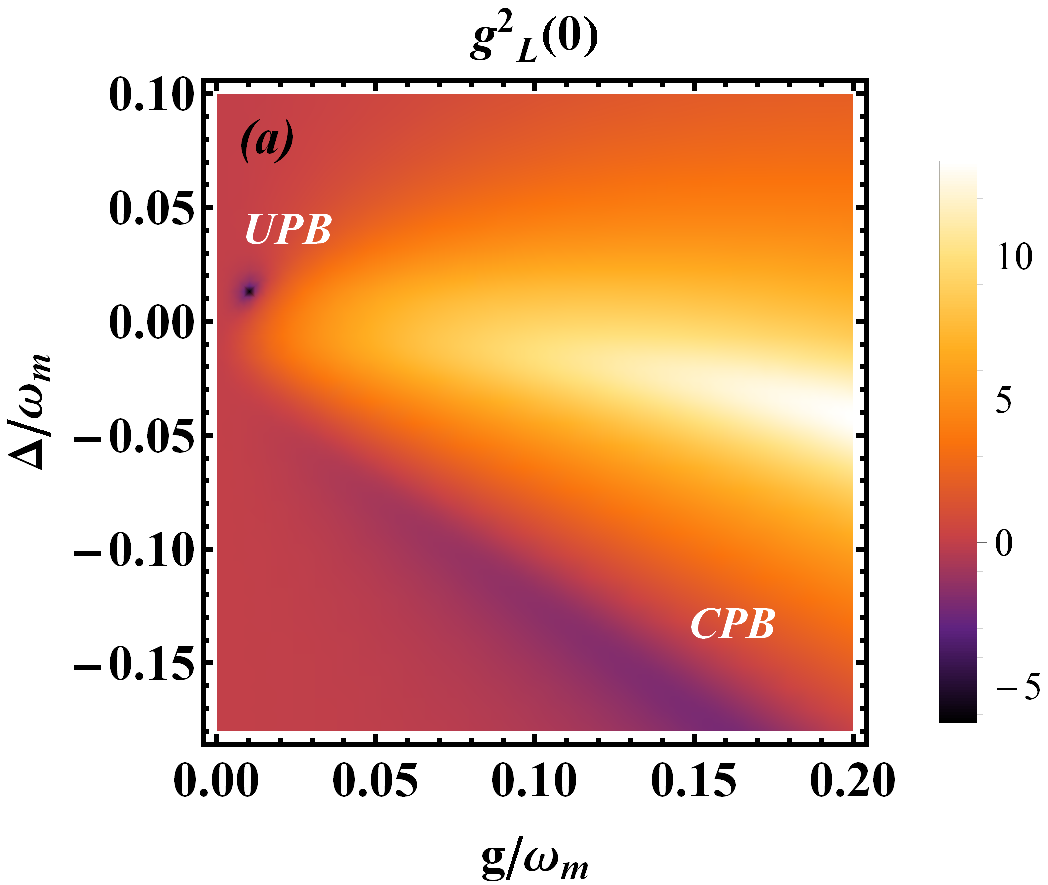} %
\includegraphics[width=4.2cm]{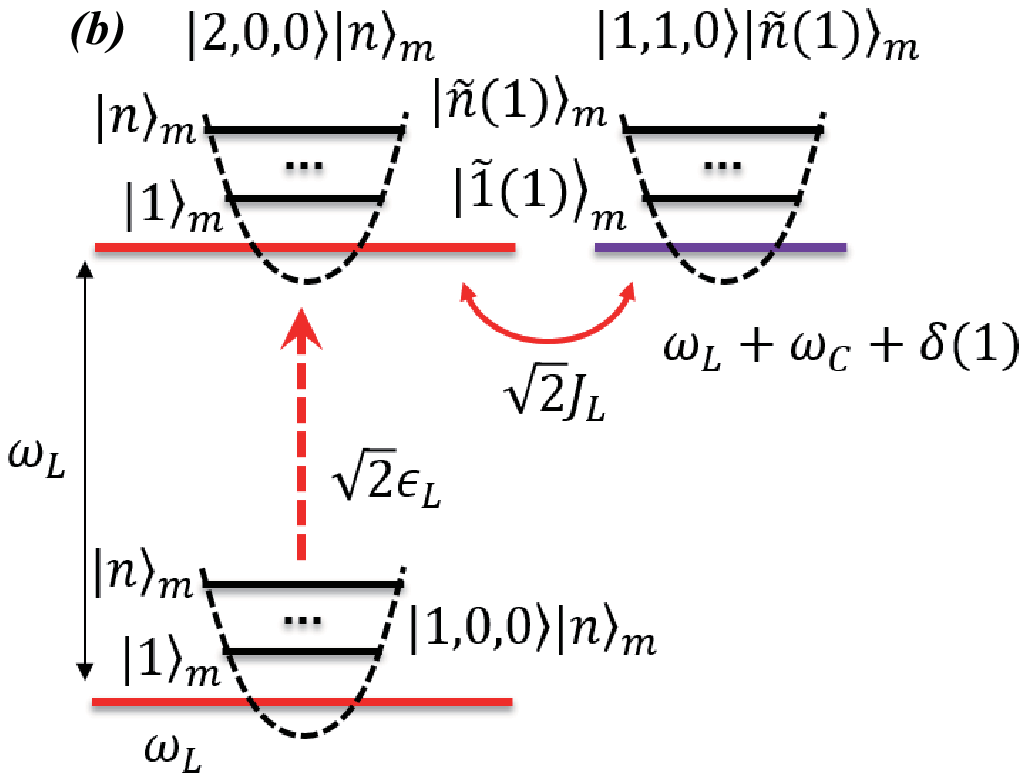}\newline
\caption{(Color online) (a)The zero-time second-order correlation $%
g_{L}^{(2)}(0)$ as a function of the coupling strength $g$ and the driving
detuning $\Delta $ (log scale). Other parameters are $J_{L}/\protect\omega %
_{m}=0.5$, $J_{R}/\protect\omega _{m}=0.01$, $\protect\kappa /\protect\omega %
_{m}=0.036$. (b) The eigensystem of two excitation path interference in
cavity $L$.}
\label{blockade}
\end{figure}

And the effect of quantum nonlinear features can be characterized by the
second-order correlation function with zero-time delay.
\begin{equation}
g_{j}^{(2)}(0)=\frac{\langle a_{j}^{\dag 2}a_{j}^{2}\rangle }{\langle
a_{j}^{\dag }a_{j}\rangle ^{2}},j=L,R,C.  \label{g20}
\end{equation}%
We notice that $g_{j}^{(2)}(0)<1$ indicates photon antibunching and $%
g_{j}^{(2)}(0)>1$ indicates photon bunching, respectively. Antibunching
corresponds to a reduced probability of two photons in the cavity at a given
time, which is the opposite for bunching. The probability of two photons in
the cavity will equal to zero if $g_{j}^{(2)}(0)\approx 0$ (photon
blockade). For simplify, we set $\kappa _{L}=\kappa _{R}=\kappa _{C}=\kappa $%
, $\alpha _{j}=\Delta _{j}-i\kappa /2$ $(j=L,C,R)$. If $\alpha _{L}=\alpha
_{R}=\alpha ,\varepsilon _{L}=\varepsilon $ and $\varepsilon
_{C},\varepsilon _{R}=0$, i.e., the system is only pumped on the left cavity
with the magnitude $\varepsilon $, the photon second-order correlation
function with no time-delay in left cavity can be obtained
\begin{eqnarray}
g_{L}^{(2)}(0) &=&|\frac{J_{R}^{4}K_{2}+(J_{L}^{2}-J_{R}^{2})K_{1}F_{2}}{%
(J_{L}^{2}+J_{R}^{2})K_{2}-K_{1}F_{2}}-F_{1}|^{2}  \nonumber \\
&&\times |\frac{\left( J_{L}^{2}+J_{R}^{2}-F_{1}\right) }{\left(
J_{R}^{2}-F_{1}\right) {}^{2}}|^{2},  \label{g2L}
\end{eqnarray}%
where $K_{n}=\alpha +\alpha _{C}+n\delta $ and $F_{n}=\alpha (\alpha
_{C}+n\delta )$ ($n=1,2$). Setting $\Delta _{L}=\Delta _{R}=\Delta
_{C}=\Delta $, we plot logarithmic $g_{L}^{(2)}(0)$ as a function of $\Delta
$ and $g$ in Fig. 3a. $g_{L}^{(2)}(0)\approx 0$ represents photon blockade,
corresponding to the dark areas, which appears in two areas. The one is
achieved in the low-right area in Fig. 3a with large values of coupling rate
$g$, which means that the photon blockade is resulted from the nonlinear
effects of radiation pressure. We call it conventional photon blockade
(CPB). And the other appears in up-left with small values of $g$ but strict
limitations on other parameters, which means that it is resulted from the
two-path interference. The interference between the two excitation path (the
one exciton is from its own exciton, and the other is the jumping from its
neighbor) is illustrated in Fig. \ref{blockade}b. Because of the destructive
interference, the photon blockade phenomenon is also appearance, called
unconventional photon blockade (UPB).

\begin{figure}[tbp]
\centering \includegraphics[width=8.5cm]{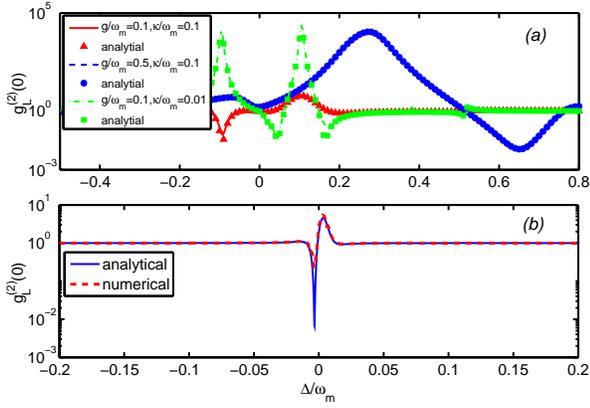}\newline
\caption{(Color online) The equal-time second-order correlation $%
g_{L}^{(2)}(0)$ as a function of the coupling strength $g$ and the driving
detuning $\Delta $, set $\Delta _{L}=\Delta _{C}=\Delta $, $\Delta
_{R}=\Delta +\Delta _{LR}$. Other parameters are (a) $J_{L}/\protect\omega %
_{m}=0.5$, $J_{R}/\protect\omega _{m}=0.1$, $\Delta _{LR}=0$. (b) $J_{L}/%
\protect\omega _{m}=0.1$, $J_{R}/\protect\omega _{m}=0.01$, $g/\protect%
\omega _{m}=0.01$, $\protect\kappa /g=1.3$, $\Delta _{LR}=0.1$ (other groups
of parameters can be get by solving $g_{L}^{(2)}(0)=0$ in weak coupling
regime).}
\label{fg2L}
\end{figure}
We now show the photon statistics properties and the controlling of the
photon transport by comparing the analytical solution with the numerical
results via solving the master equation (\ref{Meq})\textbf{.} For the
conventional photon blockade, the larger the ratio of $g/\kappa $, the
stronger the effect of blockade, shown in Fig. \ref{fg2L}a. The
corresponding detuning frequency can be derived from Eq. (\ref{g2L}). As
shown in Fig. \ref{fg2L}b, the strong photon antibunching can be obtained
even if $g/\kappa <1$.

In order to describe the characteristics of unidirectional energy transport.
We define the rectifying factor $\mathcal{R}$ and transport efficiency $%
\mathcal{T}$ as the normalized difference between the output currents when
the system is pumped through the left and right resonator (indicated by the
wave vectors $k$ and $-k$, respectively) \cite{OPD}
\begin{eqnarray}
\mathcal{R} &=&\frac{Q_{R}[k]-Q_{L}[-k]}{Q_{R}[k]+Q_{L}[-k]}, \\
\mathcal{T}_{L} &=&\frac{Q_{R}[k]}{Q_{R}[k]+Q_{L}[k]}, \\
\mathcal{T}_{R} &=&\frac{Q_{L}[-k]}{Q_{R}[-k]+Q_{L}[-k]}.
\end{eqnarray}%
$\mathcal{R}=-1$ indicates maximal rectification with enhanced transport to
the left (left rectification), $\mathcal{R}=0$ indicates no rectification
because $Q_R[+k]=Q_L[-k]$, while $\mathcal{R}=+1$ indicates maximal
rectification with transport to the right (right rectification). In our
system, cavity $L$ and cavity $R$ are both linear cavity. Therefore, there
is no rectification ($\mathcal{R}=0$) when only driving the left or right
cavity (no asymmetric nonlinear effect).

\begin{figure}[tbp]
\centering \includegraphics[width=8.5cm]{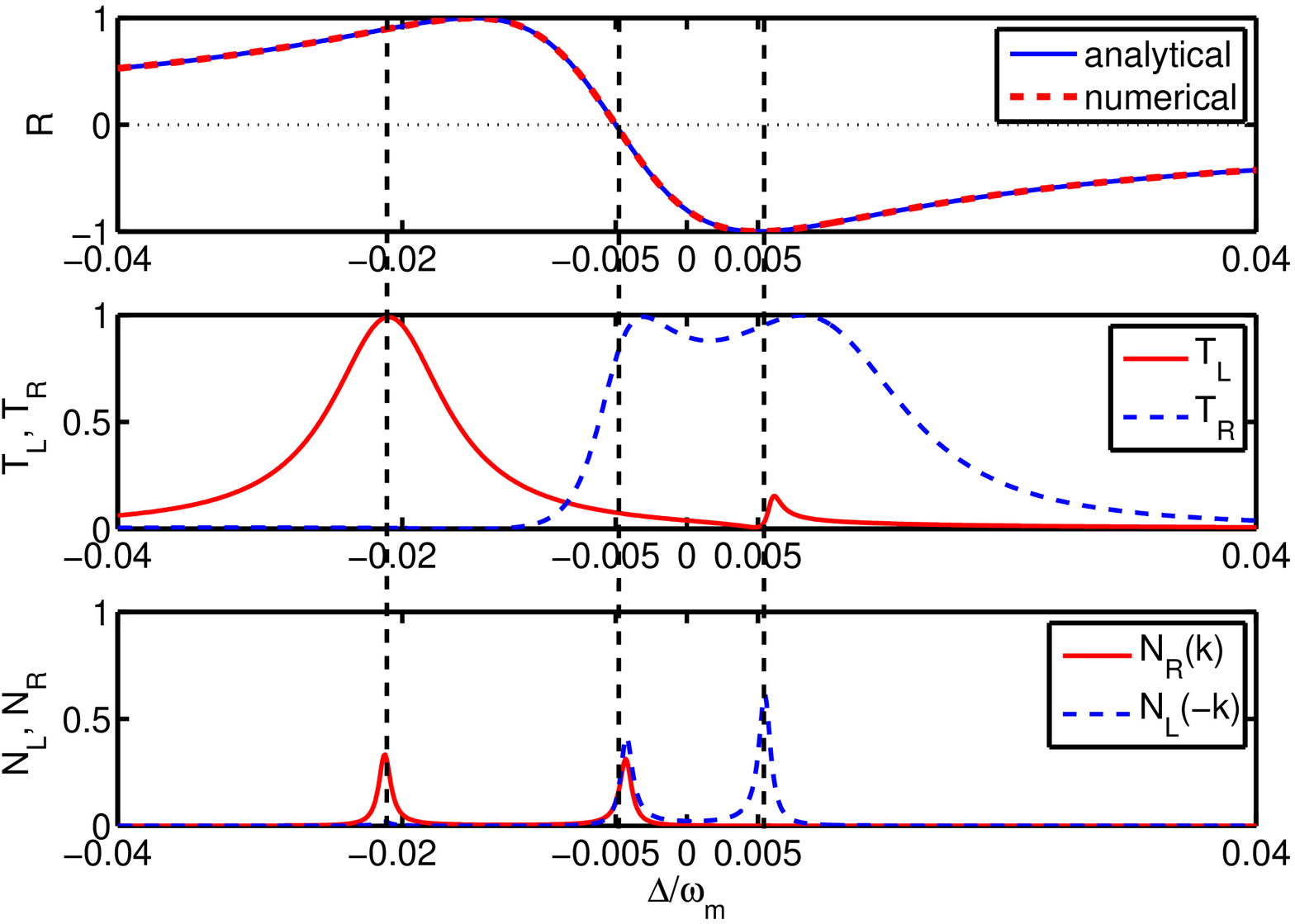}\newline
\caption{(Color online) The rectifying factor, transport efficiency and
excitation number in cavity as a function of the driving detuning $\Delta $,
here we set $\Delta _{L}=\Delta _{C}=\Delta$, $\Delta _{R}=\Delta +\Delta
_{LR}$. Other parameters are $g/\protect\omega _{m}=5\times 10^{-3}$, $J_{L}/%
\protect\omega _{m}=5\times 10^{-3}$, $J_{R}/\protect\omega _{m}=5\times
10^{-3}$, $\protect\kappa /g=0.2$, $\Delta_{LR}/\protect\omega _{m}=2\times
10^{-2}$. }
\label{fgR}
\end{figure}

We discuss the rectification effect in conventional blockade regime ($%
\omega_m>g>\kappa $) shown in Fig. \ref{fgR}. When $\frac{\Delta }{\omega
_{m}}$ is around $-0.02$, $\mathcal{R\approx }+1$ which indicates that the
system allows photon transfer from left to right $L\rightarrow R$ only, and
the transfer is prohibited from right to left $R\rightarrow L$, the photon
number from left-going ($-k$) field equal to zero in side $L$. Similarly,
when $\frac{\Delta }{\omega _{m}}$ is around 0.005, $\mathcal{R}=-1$, which
only allows the transport from right to left $R\rightarrow L$ and $%
N_{R}(k)=0 $. For $\frac{\Delta }{\omega _{m}}$ is around -0.005, $\mathcal{R%
}=0$, the photon number from left-going ($-k$) field equal to the photon
number from right-going ($k$) field $N_{L}(-k)=N_{R}(k)$. If $%
g<\kappa<\omega _{m}$ (unconventional blockade regime), shown in Fig. \ref%
{afgR}, one can obtain $\mathcal{R}\approx+1$ when $\frac{\Delta }{\omega
_{m}}=0.022$. We also can see $\mathcal{R}=0$ for $\frac{\Delta }{\omega _{m}%
}= 0.015$, and $\mathcal{R}\approx-1$ for $\frac{\Delta }{\omega _{m}}= 0.003
$. Therefore we can conclude that no matter $\omega _{m}>g>\kappa$ or $%
g<\kappa<\omega _{m}$ by tuning the frequencies of the cavity $R$ and $L$,
one can adjust (or switch) rectification and two way transport.

\begin{figure}[tbp]
\centering \includegraphics[width=8.5cm]{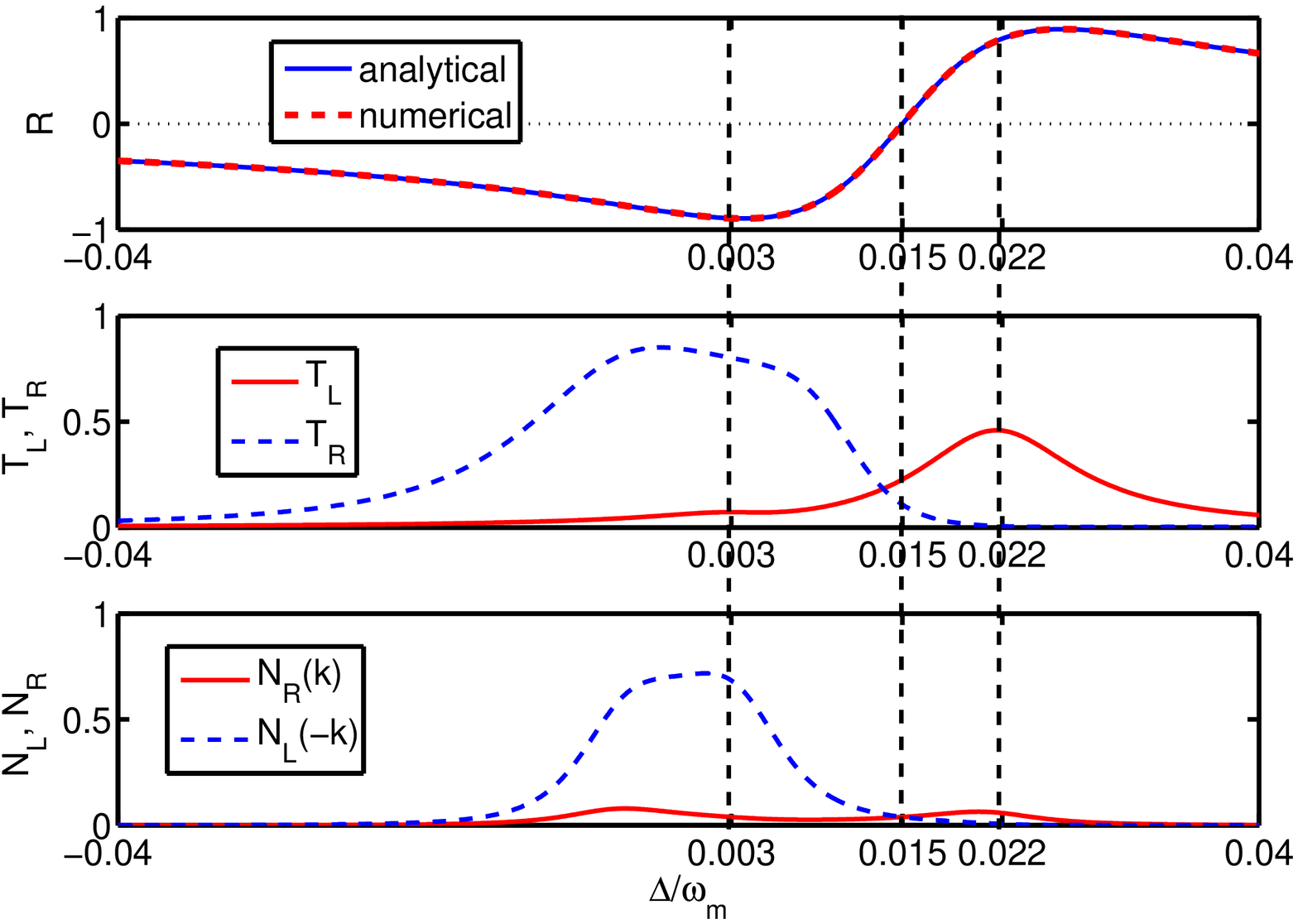}\newline
\caption{(Color online) The rectifying factor, transport efficiency and
excitation number in cavity as a function of the driving detuning $\Delta $,
here we set $\Delta _{L}=\Delta _{C}=\Delta$, $\Delta _{R}=\Delta +\Delta
_{LR}$. Other parameters are $g/\protect\omega _{m}=5\times 10^{-3}$, $%
\protect\kappa /g=2$, $J_{L}/\protect\omega _{m}=5\times 10^{-3}$, $J_{R}/%
\protect\omega _{m}=5\times 10^{-3}$, $\Delta_{LR}/\protect\omega %
_{m}=-2\times 10^{-2}$.}
\label{afgR}
\end{figure}

\subsection{Single-photon source}

Photon blockade effect allows only single-photon transmission through the
system. Now, we show that our device can work as single-photon sources. The
system is only driven from left or right cavity, i.e., $\varepsilon
_{L}=\varepsilon $, $\varepsilon _{R}=\varepsilon _{C}=0$ or $\varepsilon
_{R}=\varepsilon $, $\varepsilon _{L}=\varepsilon _{C}=0$, the mean
occupation photon numbers
\begin{eqnarray}
N_{R}(k) &=&N_{L}(-k) \\
&=&|\frac{J_{L}J_{R}\varepsilon }{J_{R}^{2}\alpha _{L}+\alpha
_{R}(J_{L}^{2}-\alpha _{L}(\alpha _{C}+\delta ))}|^{2}.  \nonumber
\end{eqnarray}

\begin{figure}[tbp]
\centering\includegraphics[width=8.5cm]{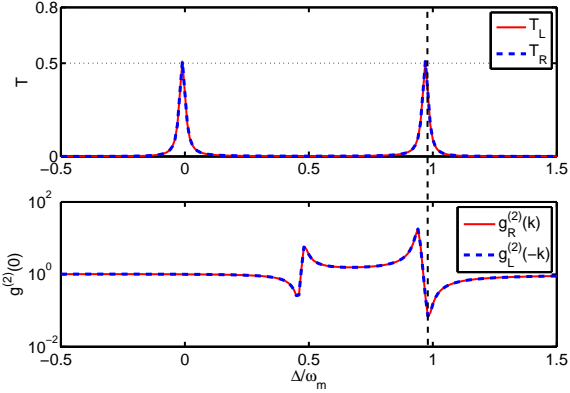}\newline
\caption{(Color online) The transport efficiency and second-order
correlation of output ports as a function of the driving detuning $\Delta $,
here we set $\Delta _{L}=\Delta _{R}=\Delta$, $\Delta _{C}=\Delta +\Delta
_{LC}$. Other parameters are $g/\protect\omega _{m}=0.2$, $J_{L}/\protect%
\omega _{m}=0.1$, $J_{R}/\protect\omega _{m}=0.1$, $\protect\kappa /g=0.1$, $%
\Delta _{LC}/\protect\omega _{m}=-1$.}
\label{ST}
\end{figure}

As shown in Fig.\ref{ST}, the system only allowed single-photon transport no
matter the light from left or right when $g_{R}^{(2)}(k)\approx
g_{L}^{(2)}(-k)\approx 0$. The transport efficiency $\mathcal{T}_{L}=%
\mathcal{T}_{R}\approx 0.5$, which means the output of system is
single-photon state if the input is two-photon state. Under this condition,
the device can control the single-photon transport in the channel or worked
as a single-photon sources. This kind of device can only worked in strong
coupling regime ($g/\kappa >1$) because there is no multi-path interference
in output ports. We also notice that, $\mathcal{R}\equiv 0$ even if $\alpha
_{L}\neq \alpha _{R},J_{L}\neq J_{R}$. In order to achieve rectification,
requires $\varepsilon _{C}\neq 0$.

\subsection{Optomechanical optical capacitor}

As we have shown in Fig. \ref{fg2L}, the photons in left cavity exhibit
antibunching although the directly nonlinear interaction only appear in OM
cavity. Similar to the nonlinear shift $3\delta$ in OM cavity, the effective
nonlinearity in cavity $L$ and $R$ can be equivalent to the resonance energy
shift $\delta_L$ ($\delta_R$). If cavity $L$ and $R$ both appear photon
antibunching effect due to the nonlinear shift and interference while the OM
cavity appears photon bunching effect, when we drive the cavity $L$ and $R$,
the photons can be stored in OM cavity. Reversing the process, one can
release the photons.

\begin{figure}[tbp]
\centering \includegraphics[width=8cm]{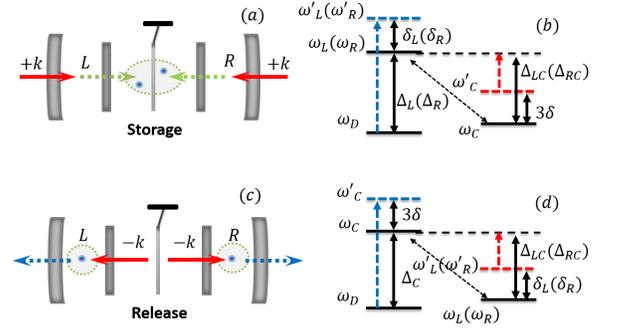}\newline
\caption{(Color Online) Pictorial representation and eigensystem of photon
storage and release progress.}
\label{system2}
\end{figure}

As shown in Fig. \ref{system2}(a) and (c), the system is a symmetric
structure. The field from left and right cavity can be regarded as input ($%
+k $) of OM, meanwhile the field from OM cavity can be regarded as output ($%
-k$) of OM. In Fig. \ref{system2}(b), when $\{\omega_L,\omega_R\}>\omega_D$
and $\{\omega_L,\omega_R\}>\omega_C$, i.e. $\{\Delta_{LC},\Delta_{RC}\}<0$, $%
\{\Delta_L,\Delta_R \}>0$, the nonlinear frequency shift in left (right)
cavity $\delta_L(\delta_R)$ will enlarge the transition energy of two-photon
excitation, which means the probability of two-photon state will be
suppressed, the photon appears anti-bunching in cavity $L$ ($R$). At the
same time, the nonlinear shift in OM cavity $3\delta$ will diminish the
detuning between tunneling field $\omega_L$ ($\omega_R$) and resonance
frequency $\omega_C$, the photon appears bunching in OM cavity. Especially
for $\omega_D\approx\omega_C+3\delta$, OM cavity exhibit strong bunching due
to the resonance absorption. Under this condition, the probability amplitude
of photons in cavity OM will be much larger than in cavity $L$ and $R$,
photons can be stored in OM cavity. Reversing the process, as shown in Fig. %
\ref{system2}(d), when $\omega_C>\omega_D$ and $\omega_C>\{\omega_L,\omega_R%
\}$, i.e. $\{\Delta_{LC},\Delta_{RC}\}>0$, $\Delta_C>0$, the nonlinear
frequency shift $3 \delta$ will enlarge the transition energy of two-photon
excitation, the photon appears antibunching in OM cavity. Meanwhile, the
nonlinear shift in left and right cavity $\delta_L$ and $\delta_R$ will
diminish the detuning between tunneling field $\omega_C$ and resonance
frequency $\{\omega_L,\omega_R\}$, the photon appears bunching in left and
right cavity. Under this condition, the photons can be released from OM
cavity.

\begin{figure}[tbph]
\centering \includegraphics[width=8.5cm]{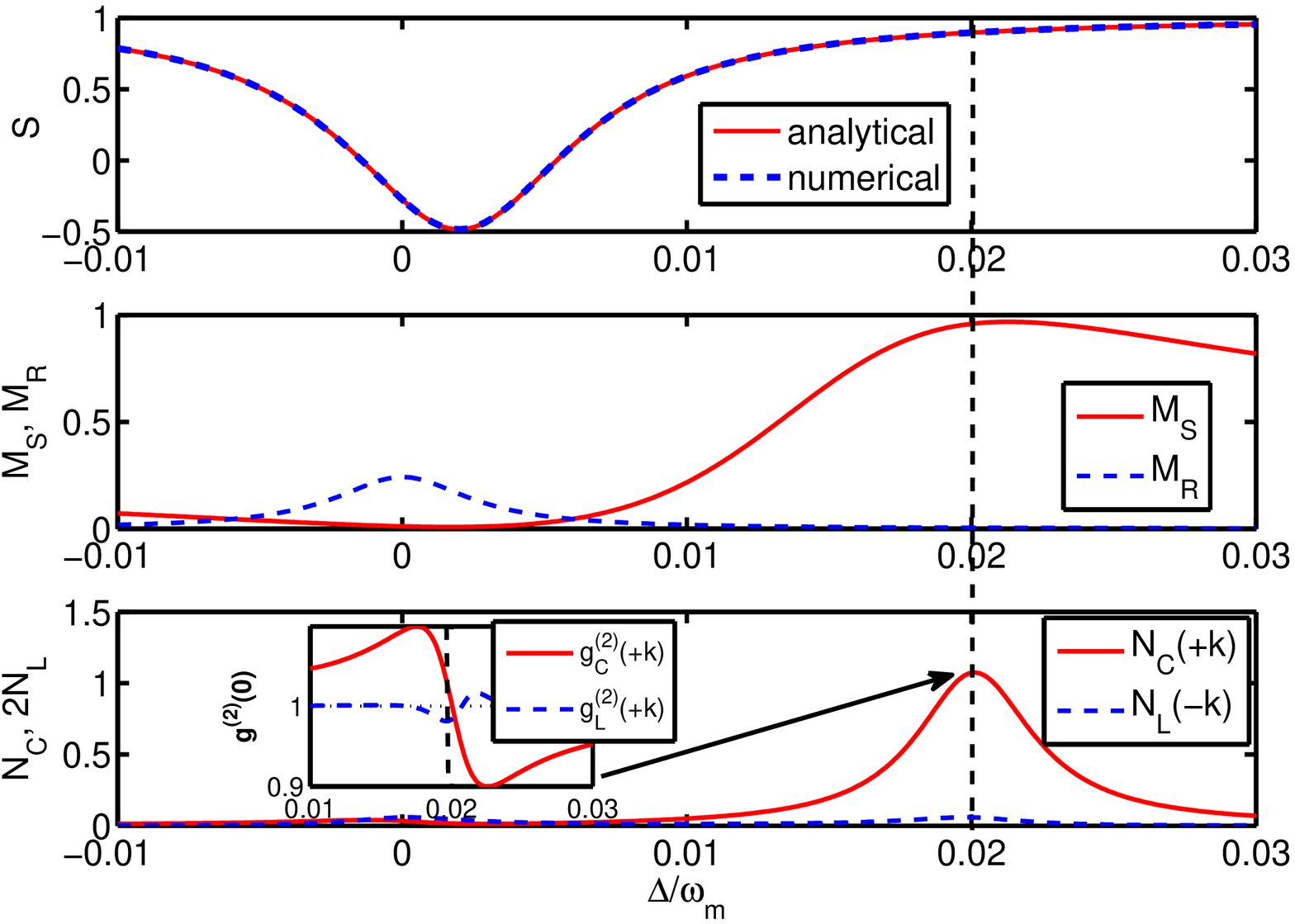}\newline
\caption{(Color online) The storage factor, storage-release efficiency and
excitation number in cavity as a function of the driving detuning $\Delta $,
here we set $\Delta _{L}=\Delta _{R}=\Delta$, $\Delta _{C}=\Delta +\Delta
_{LC} $. Other parameters are $g/\protect\omega _{m}=5\times10^{-3}$, $%
\protect\kappa /g=1$, $J_{L}/\protect\omega _{m}=J_{R}/\protect\omega %
_{m}=1\times10^{-3}$, $\Delta _{LC}/\protect\omega _{m}=-2\times10^{-2}$.}
\label{NS}
\end{figure}

In order to describe the characteristics of energy storage and release. We
define the storage factor $\mathcal{S}$ and storage-release efficiency $%
\mathcal{M}$.
\begin{eqnarray}
\mathcal{S} &=&\frac{Q_{C}[+k]-Q_{L}[-k]-Q_{R}[-k]}{%
Q_{C}[+k]+Q_{L}[-k]+Q_{R}[-k]}, \\
\mathcal{M}_{S} &=&\frac{Q_{C}[+k]}{Q_{C}[+k]+Q_{L}[+k]+Q_{R}[+k]}, \\
\mathcal{M}_{R} &=&\frac{Q_{L}[-k]+Q_{R}[-k]}{Q_{C}[-k]+Q_{L}[-k]+Q_{R}[-k]},
\end{eqnarray}%
where $\mathcal{S}=+1$ indicates maximal storage with the enhanced transport
to the OM cavity, $\mathcal{S}=0$ indicates no storage and release, while $%
\mathcal{S}=-1$ indicates maximal release with enhanced transport to the
left and right cavities. $\mathcal{M}_{S}=1$ or $\mathcal{M}_{R}=1$
indicates the photons are totally stored in or released from OM cavity,
respectively.

\begin{figure}[tbp]
\centering \includegraphics[width=8.5cm]{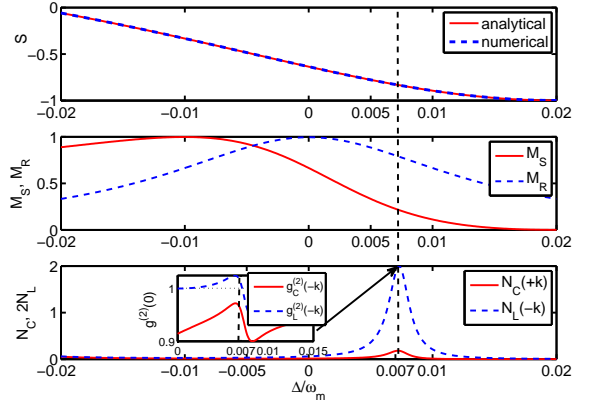}\newline
\caption{(Color online) The storage factor, storage-release efficiency and
excitation number in cavity as a function of the driving detuning $\Delta$,
here we set $\Delta_L=\Delta_R=\Delta$, $\Delta_C=\Delta+\Delta_{LC}$. Other
parameters are $g/\protect\omega_m=1\times 10^{-3}$, $\protect\kappa/g=2$, $%
J_L/\protect\omega_m=J_R/\protect\omega_m=1\times 10^{-2}$, $\Delta_{LC}/%
\protect\omega_m=2\times 10^{-2}$.}
\label{AS}
\end{figure}

For simplicity, assume that all parameters of cavity $L$ and $R$ are exactly
the same in the following discussion. The photon number of the two cavity $%
Q_{L}[+k]=Q_{R}[+k]$ and $Q_{L}[-k]=Q_{R}[-k]$. We discuss the storage
effect shown in Fig. \ref{NS}. When $\frac{\Delta}{\omega_m}$ is around $%
0.02 $, $S\approx+1$ which indicates that system allows photon transfer into
the OM cavity $L\rightarrow C\leftarrow R$ only, and the transfer is
prohibited out from the OM cavity $L\leftarrow C\rightarrow R$, the photon
number in cavity $L$ and $R$ from OM cavity approximately equal to zero. At
this time, second-order correlation $g_C^{(2)}(0)>1$, photons appear
bunching effect in OM cavity, and $g_R^{(2)}(0)=g_L^{(2)}(0)<1$, photons
appear antibunching effect in left and right cavity. The convergence filed ($%
+k$) is bounded in OM cavity, system exhibits storage characteristic. As
shown in Fig. \ref{AS}, system exhibits release characteristic. When $\frac{%
\Delta}{\omega_m}$ is around $0.007$, $S\approx-1$ which indicates that
system allows photon transfer out from OM cavity $L\leftarrow C\rightarrow R$
only. The second-order correlation $g_C^{(2)}(0)<1$, photons appear
antibunching effect in OM cavity, and $g_R^{(2)}(0)=g_L^{(2)}(0)>1$, photons
appear bunching effect in left and right cavity. That is, divergent filed ($%
-k$) is released from OM cavity. We also notice that, when $\mathcal{S}=1$
and $\mathcal{M}_{S}=1$, indicates complete storage, no matter the field
from left or right cavity can be stored in OM cavity, which is similar to
capacitor charge process. While, when $\mathcal{S}=-1$ and $\mathcal{M}%
_{R}=1 $ indicates complete release, the field in OM cavity can be released
through the left and right cavity completely, which is similar to capacitor
discharge process. And the two progress can be controlled by the detuning of
driving field $\Delta $. On the other hand, like filter effect, there is no
photon in the channel at the frequency which let $\mathcal{S}=1$ and $%
\mathcal{M}_{S}=1$ (complete absorption), but have no effect of the
frequency which let $\mathcal{S}=-1$ and $\mathcal{M}_{R}=1$ (complete
release).

\begin{figure}[tbp]
\centering \includegraphics[width=4.25cm]{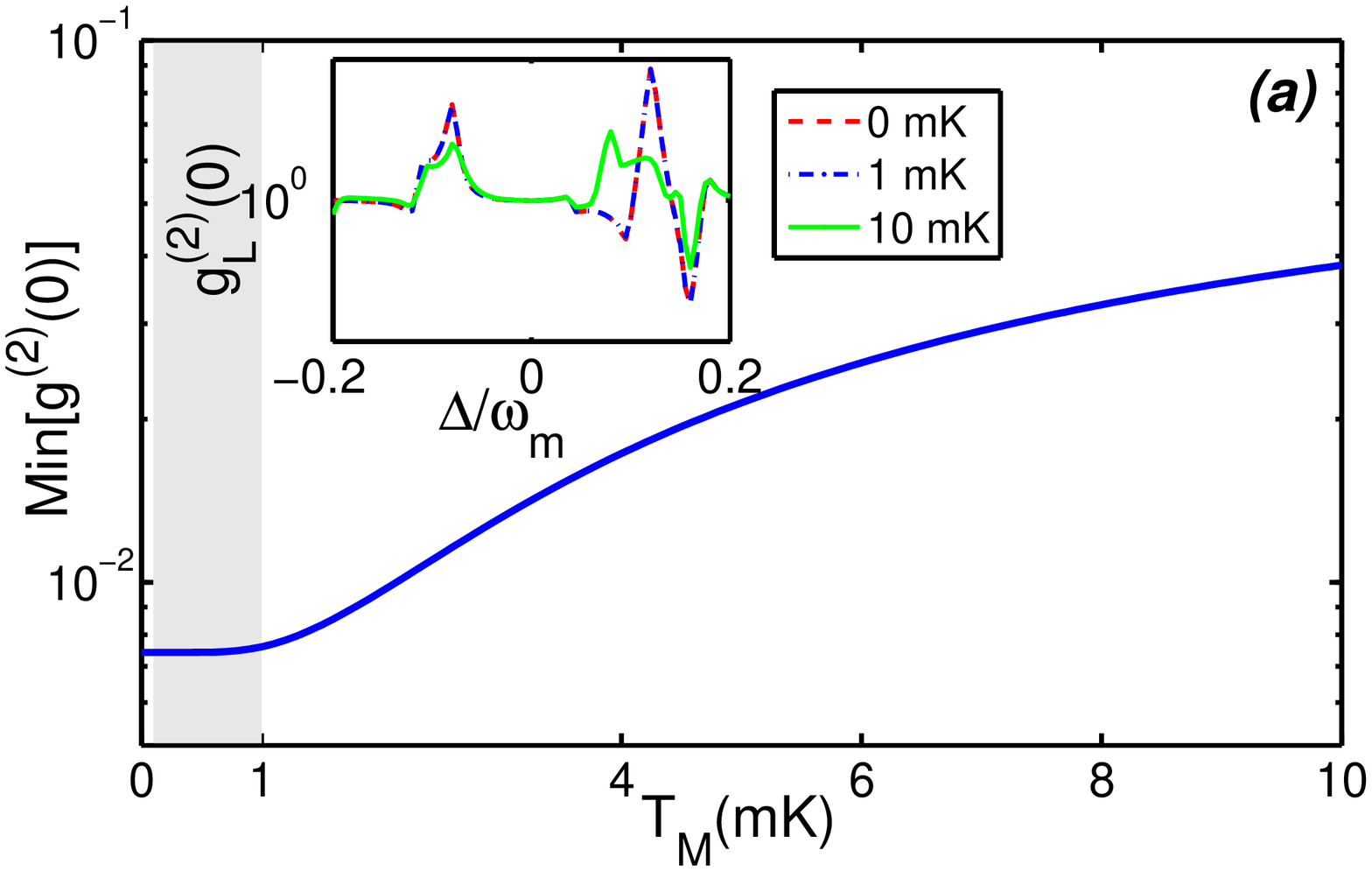} %
\includegraphics[width=4.25cm]{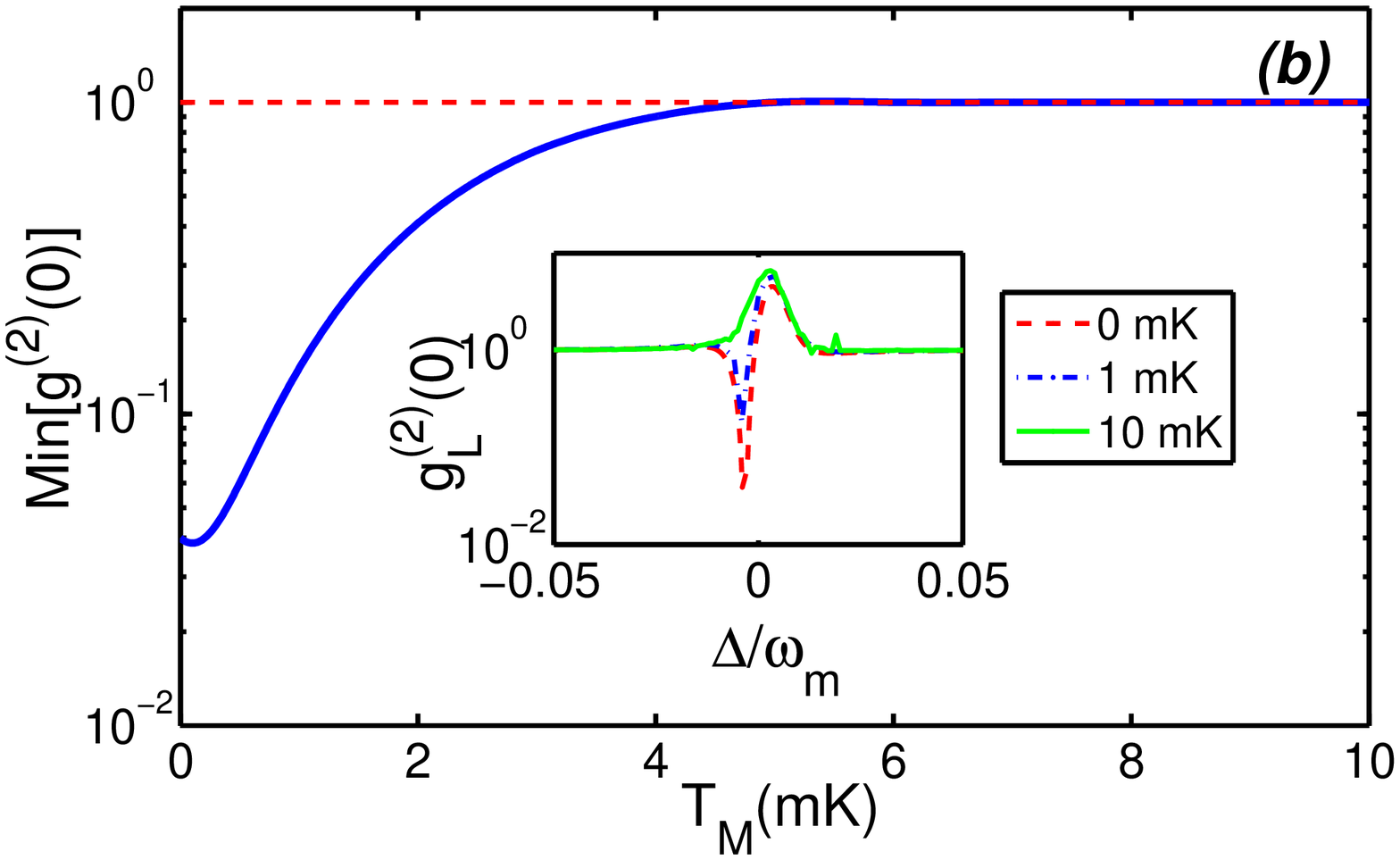}\newline
\caption{(Color online) The minimum equal-time second-order correlation $%
g_{L}^{(2)}(0)$ as a function of mechanical bath temperature $T_{M}$ with
mechanical frequency $\protect\omega _{m}=0.1GHz$ and dissipation rate $%
\protect\gamma =1KHz$. Other parameters are (a) $\Delta _{L}=\Delta
_{R}=\Delta _{C}$, $g/\protect\omega _{m}=0.2$, $J_{L}=J_{R}=\protect\omega %
_{m}/10$, $\protect\kappa /\protect\omega _{m}=10^{-2}$. (b) $\Delta
_{L}=\Delta _{C}$, $\Delta _{LR}/\protect\omega _{m}=0.1$, $g/\protect\omega %
_{m}=10^{-2}$, $J_{L}/\protect\omega _{m}=0.1$, $J_{R}/\protect\omega %
_{m}=0.01$, $\protect\kappa /g=1.3$. }
\label{TM}
\end{figure}

In the previous discussion, we ignore the effects of the mechanical thermal
bath. Now, to investigate the influence of the mechanical thermal
temperature on the correlation function, we include the mechanical thermal
reservoir. Using master Eq.(4), in Fig. 11, we plot the minimum values of $%
g_{L}^{2}(0)$ as a function of the reservoir temperature, and the $%
g_{L}^{2}(0)$ versus $\frac{\Delta }{\omega _{m}}$ affected by the thermal
reservoir are also displayed in the inset. When the temperature below $1$ $mk$
(marked with the shadow area), the thermal heating nearly have no effects in Fig. \ref{TM}(a), because when the influence of the mechanical
bath far below single-photon coupling rate, i.e. $\gamma n_{th}\ll g$, the
bath effect can be ignored. With current experimental techniques, one can
easily set $g/\gamma \gg 1$ \cite{EP1,EP2}, which means that a small value
of phonon number $n_{th}$ can be tolerance with no much effects. Also, we
can clearly see that the antibunching effect becomes more and more weaker
with temperature increasing. In UPB regime, as shown in Fig. \ref{TM}(b),
the antibunching effect is more sensitive to the bath temperature. And this
quantum effect will disappear when the temperature over $5$ $mK$ ($%
n_{th}=0.62$). Fortunately, the current experiment conditions of ground
state cooling can achieve $n_{th}=0.34\pm 0.05$ \cite{FE1}. This provides
some ability to against the quantum decoherence of our system. Even so, to
maintain the antibunching effect, the mechanical thermal noise still needs to be
suppressed.

\section{conclusion}

In this paper, we employ the radiation pressure and the destructive
interference effects to construct the controller of photon transport. By
coupling the an cavity optomechanical system to two cavities, we show that
the photon blockade can be achieved both in strong and weak coupling regime.
In strong coupling regime, the photon blockade effects is mainly resulted
from the nonlinearity of radiation pressure in optomechanical cavity, while
in weak coupling regime, the photon blockade effects is mainly because of
the interference between multipath for two-photon excitation in cavities.
For few photon control of one-dimensional transmission, the system can
worked as optical diode without the requirement of the strength of radiation
pressure strong coupling, and the rectification of photons can be controlled
by the detuning of driving field $\Delta$. If we just drive the cavity from
left or right cavity only, the system can function as single-photon source.
Furthermore, when two fields transport into the OM cavity through the cavity
$L$ and $R$, the device can store and release photons as a capacitor in
appropriate parameter regime. These novel properties provide a promising
application of optomechanical system in quantum information processing and
quantum circuit realization.

Acknowledgments: The project was supported by NSFC under Grant No. 11074028
\begin{widetext}
\appendix

\section{solution of the probability amplitudes}
We set $\kappa_L=\kappa_R=\kappa_C=\kappa$, $\omega_m\gg g$, adiabatically eliminate the degree of the oscillators, we can obtain the equations of motion of the probability amplitudes under weak pumping regime $C_{0}\gg C_{s1}\gg C_{s1s2}$. We drop higher-order terms in the zero- and one-photon probability amplitudes
\begin{eqnarray}
i\dot{C_L} &=& \alpha_L C_L+\varepsilon_L C_0+J_L C_C, \nonumber\\
i\dot{C_C} &=& (\alpha_C+\delta)C_C+\varepsilon_C C_0+J_L C_L+J_R C_R,\nonumber\\
i\dot{C_R} &=& \alpha_R C_R+\varepsilon_R C_0+J_R C_C, \nonumber\\
i\dot{C_{LC}}&=& (\alpha_L+\alpha_C+\delta)C_{LC}+\varepsilon_L C_C+\varepsilon_C C_L+\sqrt{2}J_L(C_{LL}+C_{CC})+J_RC_{LR},\nonumber\\
i\dot{C_{LR}}&=&(\alpha_L+\alpha_R)C_{LR}+\varepsilon_L C_R+\varepsilon_R C_L+J_L C_{CR}+J_R C_{LC},\nonumber\\
i\dot{C_{CR}}&=& (\alpha_R+\alpha_C+\delta)C_{CR}+\varepsilon_C C_R+\varepsilon_R C_C+\sqrt{2}J_R(C_{RR}+C_{CC})+J_L C_{LR},\nonumber\\
i\dot{C_{LL}}&=&2 \alpha_L C_{LL}+\sqrt{2} \varepsilon_L C_L+\sqrt{2} J_L C_{LC},\nonumber\\
i\dot{C_{CC}}&=& (2\alpha_C+4\delta)C_{CC}+\sqrt{2} \varepsilon_C C_C+\sqrt{2}J_L C_{LC}+\sqrt{2}J_R C_{CR},\nonumber\\
i\dot{C_{RR}}&=& 2 \alpha_R C_{RR}+\sqrt{2} \varepsilon_R C_R+\sqrt{2} J_R C_{CR},\label{effeqs}
\end{eqnarray}
where $\alpha_L=\Delta_L-i \kappa/2$, $\alpha_R=\Delta_R-i\kappa/2$, $\alpha_C=\Delta_C-\kappa/2+\delta$. If we set the initial state is vacuum state, i.e. $%
C_{0}(0)=1,C_{s1}(0)=C_{s1s2}(0)=0,\{ s1,s2\}\in\{ L,C,R\}$, In the weak-driving regime, $\{\varepsilon _{C}/\kappa _{C}$,$\varepsilon
_{L}/\kappa _{L}$,$\varepsilon _{R}/\kappa _{R}\}\ll 1$, the photon number
is small, so we have $C_0(\infty)\approx C_0(0)$, then the long-time solution of equations can be approximately obtained as,
\begin{eqnarray}
  C_L &=& \frac{[-J_R^2  + \alpha_R (\alpha_C + \delta)] \varepsilon_L +
 J_L (-\alpha_R \varepsilon_C + J_R \varepsilon_R)}{D_1},\\
  C_C&=&\frac{\alpha_L \alpha_R \varepsilon_C - J_L \alpha_R \varepsilon_L -
 J_R \alpha_L \varepsilon_R}{D_1},\\
  C_R&=& \frac{[-J_L^2 + \alpha_L (\alpha_C + \delta)]\varepsilon_R+J_R (-\alpha_L \varepsilon_C +J_L \varepsilon_L)}{D_1},\\
  C_{LL}&=&\frac{C_L\sum_{j=(L,C,R)}l_{L,j}\varepsilon_j+J_LC_C\sum_{j=(L,C,R)}l_{C,j}\varepsilon_j+J_LJ_RC_R\sum_{j=(L,C,R)}l_{R,j}\varepsilon_j}{\sqrt{2}D_2},\\
  C_{CC}&=&\frac{J_LC_L\sum_{j=(L,C,R)}c_{L,j}\varepsilon_j+C_C\sum_{j=(L,C,R)}c_{C,j}\varepsilon_j+J_RC_R\sum_{j=(L,C,R)}c_{R,j}\varepsilon_j}{\sqrt{2}D_2},\\
  C_{RR}&=&\frac{J_LJ_RC_L\sum_{j=(L,C,R)}r_{L,j}\varepsilon_j+J_RC_C\sum_{j=(L,C,R)}r_{C,j}\varepsilon_j+C_R\sum_{j=(L,C,R)}r_{R,j}\varepsilon_j}{\sqrt{2}D_2},
\end{eqnarray}
where the first term in Eq. (A5) describes two-photon state generated by driving field in cavity $L$, the second term describes two-photon excitation due to photon tunneling between OM cavity and left cavity with coupling rate $J_L$, the third term describes two-photon excitation due to photon tunneling between right cavity and left cavity through the OM cavity, when the collective effect of this three progress let $C_{LL}\approx0$, photons exhibit blockade effect in cavity $L$. As well as Eq. (A6) and (A7).
\begin{eqnarray}
D_1&=&J_L^2 \alpha_L+J_R^2 \alpha_R-\alpha_L \alpha_R(\alpha_C+\delta),\nonumber\\
D_2&=&\sum_{s=L,R}\alpha_s[J_s^2-\alpha_s(\alpha_s+\alpha_C+\delta)][2J_s^2(\alpha_s+\alpha_C+2\delta)-
\alpha_s(\alpha_s+\alpha_C+\delta)(\alpha_C+2\delta)],\nonumber\\
l_{L,L}&=&J_L^4 \alpha_R + [J_R^2 - (\alpha_L + \alpha_R) (\alpha_C +\alpha_L + \delta)] [-\alpha_R (\alpha_C + \alpha_R + \delta)(\alpha_C + 2 \delta) +J_R^2 (\alpha_C + \alpha_R + 2 \delta)]\nonumber\\
&&+ J_L^2 [J_R^2 (\alpha_L - \alpha_R) - \alpha_R (\alpha_C^2 +\alpha_R (\alpha_L + \alpha_R) + (3 \alpha_L + \alpha_R)\delta+2 \delta^2 + \alpha_C (2 \alpha_L + \alpha_R +3 \delta))],\nonumber\\
l_{L,C}&=&J_L [J_R^2 (\alpha_L + \alpha_R) (\alpha_C + \alpha_R +2 \delta) - \alpha_R (\alpha_C +
      2 \delta) (-J_L^2 + (\alpha_L + \alpha_R) (\alpha_C +\alpha_R + \delta))],\nonumber\\
l_{L,R}&=&J_LJ_R [-J_R^2 (\alpha_C + \alpha_R +2 \delta) + \alpha_R (J_L^2 + (\alpha_C + \alpha_R +\delta) (\alpha_C + 2 \delta))],\nonumber\\
l_{C,L}&=&-J_R^2(\alpha_L + \alpha_R) (\alpha_C + \alpha_R +2 \delta) +\alpha_R (\alpha_C +2 \delta) [-J_L^2 + (\alpha_L + \alpha_R) (\alpha_C + \alpha_R + \delta))],\nonumber\\
l_{C,C}&=&J_L[J_R^2 \alpha_L+J_L^2 \alpha_R - \alpha_R (\alpha_L + \alpha_R) (\alpha_C +\alpha_R+\delta)],\nonumber\\
l_{C,R}&=&J_LJ_R\alpha_R (\alpha_C + \alpha_L + \alpha_R + 2 \delta),\nonumber\\
l_{R,L}&=&J_R^2 (\alpha_C + \alpha_R +2 \delta) - \alpha_R [J_L^2 + (\alpha_C + \alpha_R +\delta) (\alpha_C + 2\delta)],\nonumber\\
l_{R,C}&=&J_L \alpha_R (\alpha_C + \alpha_L + \alpha_R + 2 \delta),\nonumber\\
l_{R,R}&=&-J_L J_R (\alpha_C + \alpha_L + \alpha_R + 2 \delta),\nonumber\\
c_{L,L}&=&J_L [J_R^2 \alpha_L +J_L^2 \alpha_R - \alpha_R (\alpha_L + \alpha_R) (\alpha_C +\alpha_R + \delta)],\nonumber\\
c_{L,C}&=&\alpha_L [-J_R^2 \alpha_L -J_L^2 \alpha_R + \alpha_R (\alpha_L + \alpha_R) (\alpha_C +\alpha_R + \delta)],\nonumber\\
c_{L,R}&=&J_R [J_R^2 \alpha_L - \alpha_R (-J_L^2 + \alpha_L (2 \alpha_C +\alpha_L + \alpha_R + 2 \delta))],\nonumber\\
c_{C,L}&=&J_L \alpha_L [-J_R^2 \alpha_L -J_L^2 \alpha_R + \alpha_R (\alpha_L + \alpha_R) (\alpha_C +\alpha_R + \delta)],\nonumber\\
c_{C,C}&=&-J_L^4 \alpha_R - \alpha_L [J_R^2 - (\alpha_L + \alpha_R)(\alpha_C + \alpha_L + \delta)] [J_R^2 - \alpha_R (\alpha_C +\alpha_R + \delta)]\nonumber\\
&& + J_L^2 [-J_R^2 (\alpha_L + \alpha_R) + \alpha_R (\alpha_L^2 +\alpha_C (2 \alpha_L + \alpha_R) + \alpha_R (\alpha_R +\delta) + \alpha_L (\alpha_R + 2 \delta))],\nonumber\\
c_{C,R}&=&J_R \alpha_R [-J_R^2 \alpha_L -J_L^2 \alpha_R + \alpha_L (\alpha_L +\alpha_R) (\alpha_C +\alpha_L + \delta)],\nonumber\\
c_{R,L}&=&J_L [J_R^2 \alpha_L - \alpha_R (-J_L^2 + \alpha_L (2 \alpha_C +\alpha_L + \alpha_R + 2 \delta))],\nonumber\\
c_{R,C}&=&\alpha_R [-J_R^2 \alpha_L -J_L^2 \alpha_R + \alpha_L (\alpha_L + \alpha_R) (\alpha_C +\alpha_L +\delta)],\nonumber\\
c_{R,R}&=&J_R [J_R^2 \alpha_L +J_L^2 \alpha_R - \alpha_L (\alpha_L + \alpha_R) (\alpha_C + \alpha_L + \delta)],\nonumber\\
r_{i,j}&=&l_{i,j}(J_L\leftrightarrow J_R,\alpha_L\leftrightarrow \alpha_R),(i,j=\{L,C,R\})
\end{eqnarray}
The second order correlation functions with zero time-delay are
\begin{eqnarray}
g_{L}^{(2)}(0) &=&\frac{2|C_{LL}|^{2}}{%
(|C_{L}|^{2}+|C_{LC}|^{2}+|C_{LR}|^{2}+2|C_{LL}|^{2})^{2}}\nonumber\\
&\approx &\frac{2|C_{LL}|^{2}}{|C_{L}|^{4}},\nonumber \\
g_{C}^{(2)}(0) &=&\frac{2|C_{CC}|^{2}}{%
(|C_{C}|^{2}+|C_{LC}|^{2}+|C_{CR}|^{2}+2|C_{CC}|^{2})^{2}}\nonumber\\
&\approx &\frac{2|C_{CC}|^{2}}{|C_{C}|^{4}},\nonumber \\
g_{R}^{(2)}(0) &=&\frac{2|C_{RR}|^{2}}{%
(|C_{R}|^{2}+|C_{CR}|^{2}+|C_{LR}|^{2}+2|C_{RR}|^{2})^{2}}\nonumber\\
&\approx &\frac{2|C_{RR}|^{2}}{|C_{R}|^{4}}.
\end{eqnarray}
The mean occupation numbers of three cavities are
\begin{eqnarray}
N_{L} &=&(|C_{L}|^{2}+|C_{LC}|^{2}+|C_{LR}|^{2}+2|C_{LL}|^{2})N_{0}
\nonumber \\
&\approx &|C_{L}|^{2}N_{0},  \nonumber \\
N_{C} &=&(|C_{C}|^{2}+|C_{LC}|^{2}+|C_{CR}|^{2}+2|C_{CC}|^{2})N_{0}
\nonumber \\
&\approx &|C_{C}|^{2}N_{0},  \nonumber \\
N_{R} &=&(|C_{R}|^{2}+|C_{CR}|^{2}+|C_{LR}|^{2}+2|C_{RR}|^{2})N_{0}
\nonumber \\
&\approx &|C_{R}|^{2}N_{0},
\end{eqnarray}
where $N_{0}=(\varepsilon _{L}/\kappa _{L})^{2}+(\varepsilon _{C}/\kappa
_{C})^{2}+(\varepsilon _{R}/\kappa _{R})^{2}$.

\end{widetext}

\end{document}